\documentclass[apj,square,numbers,iop]{myemulateapj}
\usepackage{graphicx}
\usepackage{xcolor}
\usepackage{amsmath,amssymb,latexsym,bm}
\usepackage{hyperref}
\usepackage{ulem}
\usepackage{times}

\begin{document}
\title{Model-independent Constraints on Cosmic Curvature and Opacity}
\author{Guo-Jian Wang$^{1}$, Jun-Jie Wei$^{2,3}$, Zheng-Xiang Li$^{1}$, Jun-Qing Xia$^{1}$, Zong-Hong Zhu$^{4,1}$}

\affil{$^{1}$Department of Astronomy, Beijing Normal University, Beijing 100875, China;\\gjwang@mail.bnu.edu.cn, zxli918@bnu.edu.cn, xiajq@bnu.edu.cn, zhuzh@bnu.edu.cn}
\affil{$^{2}$Purple Mountain Observatory, Chinese Academy of Sciences, Nanjing 210008, China; jjwei@pmo.ac.cn}
\affil{$^{3}$Guangxi Key Laboratory for Relativistic Astrophysics, Nanning 530004, China}
\affil{$^{4}$School of Physics and Technology, Wuhan University, Wuhan 430072, China}

\begin{abstract}
In this paper, we propose to estimate the spatial curvature of the universe and the cosmic opacity in a model-independent way with expansion rate measurements, $H(z)$, and type Ia supernova (SNe Ia). On the one hand, using a nonparametric smoothing method Gaussian process, we reconstruct a function $H(z)$ from opacity-free expansion rate measurements. Then, we integrate the $H(z)$ to obtain distance modulus $\mu_{\rm H}$, which is dependent on the cosmic curvature. On the other hand, distances of SNe Ia can be determined by their photometric observations and thus are opacity-dependent. In our analysis, by confronting distance moduli $\mu_{\rm H}$ with those obtained from SNe Ia, we achieve estimations for both the spatial curvature and the cosmic opacity without any assumptions for the cosmological model. Here, it should be noted that light curve fitting parameters, accounting for the distance estimation of SNe Ia, are determined in a global fit together with the cosmic opacity and spatial curvature to get rid of the dependence of these parameters on cosmology. In addition, we also investigate whether the inclusion of different priors for the present expansion rate ($H_0$: global estimation, $67.74\pm 0.46~\rm km~ s^{-1} ~Mpc^{-1}$, and local measurement, $73.24\pm 1.74~\rm km~ s^{-1} ~Mpc^{-1}$) exert influence on the reconstructed $H(z)$ and the following estimations of the spatial curvature and cosmic opacity. Results show that, in general, a spatially flat and transparent universe is preferred by the observations. Moreover, it is suggested that priors for $H_0$ matter a lot. Finally, we find that there is a strong degeneracy between the curvature and the opacity.
\end{abstract}

\maketitle

\section{Introduction}\label{sec:introduction}
The cosmic curvature is a fundamental parameter in modern cosmology. Important problems, such as the evolution of the universe, are related to the curvature of the universe. In theory, models of inflation predict that the radius of the curvature of the universe should be very large, which in turn implies a very small cosmic curvature. In practice, constraints from several popular cosmological probes, such as type Ia supernova (SNe Ia), cosmic microwave background radiation, and baryon acoustic oscillation (BAO), and expansion rate measurements have implied that the space of the universe is very flat \citep{Komatsu:2011}. More recently, the latest $Planck$ 2015 \citep{Ade:2016} has achieved a very precise estimation for the density parameter of curvature: $\Omega_K = 0.000\pm0.005$ (95\%, $Planck$ TT+lowP+lensing+BAO). Although the cosmic curvature has been constrained with great precision, it should be stressed that these estimations are based on the fact that the background of the universe is described by the homogeneous and isotropic Friedmann-Lema\^{\i}tre-Robertson-Walker (FLRW) metric. Another intractable issue is the degeneracy between the spatial curvature $\Omega_K$ and the equation of state parameter $w(z)$ of dark energy. This degeneracy makes it difficult to constrain these two parameters simultaneously, and prevents us from understanding the nature of dark energy. In the literature, a spatially flat universe ($\Omega_K$=0) is usually fixed  to study  $w(z)$. Likewise, the standard $\Lambda$ plus cold dark matter ($\Lambda$CDM) or the constant $w$ dark energy scenario ($w$CDM) is usually assumed to study the cosmic curvature. Therefore, any one of the cosmological-model-independent estimations for the cosmic curvature can be effective in breaking the degeneracy, and thus quite helpful for studying the nature of dark energy.

Recently, \citet{Clarkson:2007,Clarkson:2008} proposed a model-independent method to test the FLRW metric by combining the Hubble parameter $H(z)$ and the transverse comoving distance $D_M(z)$. An important by-product of this method is that, if the FLRW metric is valid, we can estimate the curvature of the universe with the expression
\begin{equation}\label{equ:clarkson}
\Omega_{k}=\frac{\left[H(z)D_M'(z)\right]^{2}-c^{2}}{\left[H_{0}D_M(z)\right]^{2}} ~,
\end{equation}
where $H_{0}$ is the Hubble constant, $c$ is the speed of light, $D_M(z)=(1+z)D_A(z)=D_L(z)/(1+z)$ is the  transverse comoving distance \citep{Hogg:1999}, and $``'"$ represents the derivative with respect to redshift $z$. This method has been widely used in the literature to test the FLRW metric or estimate the cosmic curvature \citep{Shafieloo:2010,Mortsell:2011,Li:2014,Sapone:2014,Yahya:2014,Cai:2016,Rana:2016}. However, in this method, the derivative of transverse comoving distance to redshift, $D_M'(z)$, which introduces a large uncertainty, is necessary to estimate the cosmic curvature. Besides, \citet{Bernstein:2006} proposed another model-independent method to constrain the cosmic curvature parameter $\Omega_K$ based on the sum rule of distances along null geodesics of the FLRW metric. More recently, this distance sum rule was put forward to test the validity of the FLRW metric \citep{Rasanen:2015}. Similarly, this method can also achieve estimation for the cosmic curvature if the validity of the FLRW metric is confirmed. In their analysis, the Union2.1 SNe Ia \citep{Suzuki:2012} and strong gravitational lensing systems selected from the Sloan Lens ACS Survey \citep{Bolton:2008} were used. However, the cosmic curvature was weakly constrained due to the large uncertainties of the gravitational lensing systems. In addition, because the light curve fitting parameters in the Union2.1 SNe Ia are determined using global fitting by assuming the standard dark energy model with the equation of state being constant, the measurement of the cosmic curvature was not completely model-independent. 

To avoid shortcomings of the methods mentioned above, on the one hand, \citet{Yu:2016} proposed to estimate the cosmic curvature by combining the proper distance, $d_P$, and the transverse comoving distance, $D_M$, using measurements from the Hubble parameter $H(z)$ and the angular diameter distance, $d_A$, of BAO for the method proposed in \citet{Clarkson:2007,Clarkson:2008}. However, $d_A$ and some of the $H(z)$ measurements used in their analysis are obtained from the BAO observations, which are dependent on the assumed fiducial cosmological model. On the other hand, \citet{Li:2016} and \citet{Wei:2017} proposed to constrain the cosmic curvature in a model-independent way by combining the $H(z)$ and SNe Ia data. They both concluded that the current observations are well consistent with a spatially flat universe.

Since the cosmic acceleration was first revealed by observations of SNe Ia \citep{Riess:1998, Perlmutter:1999}, SNe Ia plays an essential role in studying the expansion history of the universe and the nature of dark energy. However, there are still some open issues in SNe Ia cosmology. For example, \citet{Betoule:2014} found that constraints on cosmological parameters inferred from SNe Ia are limited by systematic errors rather than by statistical ones. One of the important systematic errors might come from the cosmic opacity. In recent years, the cosmic distance duality (CDD) relation has been used to test the existence of exotic physics as well as the presence of opacity and systematic errors in SNe Ia observations. The CDD relation is written as  \citep{Etherington:1993,Ellis:2007}
\begin{equation}\label{equ:CDDR}
D_L = D_A(1+z)^2 ~,
\end{equation}
where $D_L$ is the luminosity distance and $D_A$ is the angular diameter distance. So far, a great deal of effort has been made to test the validity of the CDD relation \citep{Gurvits:1994,Daly:2003,Bassett:2004,Holanda:2010,Holanda:2011,Holanda:2012a,Lv:2016,More:2016,Fu:2017,Rana:2017}. Meanwhile, assuming that the deviation of the CDD relation is caused completely by the nonconservation of photon number, the cosmic opacity has been widely studied with different kinds of observations \citep{Avgoustidis:2009,Avgoustidis:2010,Amanullah:2010,Lima:2011,Chen:2012,Gonçalves:2012,Holanda:2012b,Holanda:2013,Li:2013,Liao:2013,Liao:2015,Holanda:2014,Jesus:2016,Hu:2017}. In these works, some were carried out by assuming a fiducial cosmological model and others were carried out using cosmological-model-independent tests. However, it should be noted that the one thing that all of these works have in common is the flat FLRW background. Likewise, as we have mentioned above, all of the tests for the FLRW metric and all of the studies on the cosmic curvature with SNe Ia observations have also neglected the influence of the cosmic opacity.

In this paper, on the basis of \citet{Li:2016} and \citet{Wei:2017}, we propose to test the cosmic curvature and opacity simultaneously by confronting the measurements of the Hubble parameter $H(z)$ with SNe Ia observations. Firstly, we use a Gaussian process (GP), a nonparametric method, to reconstruct a function $H(z)$ and further obtain the corresponding distance modulus $\mu_{\rm H}$. Next, distances of SNe Ia are empirically expressed as a combination of some of the observables of them. Finally, we estimate $\Omega_K$ and $\epsilon$ by comparing $\mu_{\rm H}$ with the distance modulus of SNe Ia. In our analysis, the light curve parameters of SNe Ia are set free to dodge any dependence on the assumption for the energy-momentum content of the universe. In addition, we use two priors of $H_0$, $67.74\pm 0.46~\rm km~ s^{-1} ~Mpc^{-1}$ and $73.24\pm 1.74~\rm km~ s^{-1} ~Mpc^{-1}$ when reconstructing a function $H(z)$ to test the influence of $H_0$ on the constraint on the parameters.

This paper is organized as follows. Section \ref{sec:data_method} is dedicated to describing the data and method used in our analysis. Section \ref{sec:results} shows the numerical results. Finally, conclusions and discussions are presented
in Section \ref{sec:conclusions_discussions}.

\section{Observational data and Method}\label{sec:data_method}

In this section, we describe the data sets and method used in our analysis. The key point of our analysis is to obtain the distance modulus $\mu_{\rm H}$ from the Hubble parameter data and $\mu_{\rm SNe}$ from the SNe Ia data. Then, we constrain $\Omega_K, \epsilon$, and other nuisance parameters simultaneously by comparing $\mu_{\rm H}$ and $\mu_{\rm SNe}$.

\subsection{Hubble parameter data}\label{subsec:Hubble_param_data}

The Hubble parameter measurements $H(z)$, which describe the expansion rate of the universe, have been used extensively for the exploration of the evolution of the universe and the nature of dark energy. $H(z)$ can be obtained in two ways. One method is to calculate the derivative of the cosmic time with respect to the redshift at $z\neq0$,
\begin{equation}
H(z)\simeq-\frac{1}{1+z}\frac{\Delta z}{\Delta t} ~,
\end{equation}
as first presented by \citet{Jimenez:2002}. The key to this method consists of measuring the difference of age for two red galaxies at different redshifts in order to obtain the rate of $\Delta z/\Delta t$. In this method, the cosmic opacity is not strongly wavelength-dependent in the optical band. Hence, the $H(z)$ data are opacity-free, and it are usually called the cosmic chronometers (hereafter CC $H(z)$). The other method to obtain $H(z)$ is based on the detection of the radial BAO features \citep{Gaztanaga:2009,Blake:2012,Samushia:2013}. However, this method is obviously based on an assumed cosmological model. Hence, $H(z)$ obtained from
this method are not included in our model-independent analysis. Specifically, we select 30 CC $H(z)$ (which are model-independent) of the latest 41 $H(z)$ sample compiled in \citet{Wei:2017} to conduct our analysis. Note that the datapoint from \citet{Chuang:2012} in Table 1 of \citet{Wei:2017} is not from the cosmic chronometers, but from the large scale structure, which is model-dependent. So, here we eliminate this datapoint and, for convenience, list the 30 CC $H(z)$ data in Table \ref{tab:Hz}.

\begin{table}
\centering \caption{CC $H(z)$ measurements obtained from the differential age method.}\label{tab:Hz}
\begin{tabular}{ccc}
\hline
\hline
z 		& $H(z)$ (km $\rm s^{-1}$ $\rm Mpc^{-1}$) & References \\
\hline
0.09	&	$69\pm12$		&	\citet{Jimenez:2003} \\
\hline
0.17	&	$83\pm8$		&	\\
0.27	&	$77\pm14$		&	\\
0.4		&	$95\pm17$		&	\\
0.9		&	$117\pm23$		&	\citet{Simon:2005} \\
1.3		&	$168\pm17$		&	\\
1.43	&	$177\pm18$		&	\\
1.53	&	$140\pm14$		&	\\
1.75	&	$202\pm40$		&	\\
\hline
0.48	&	$97\pm62$		&	\citet{Stern:2010} \\
0.88	&	$90\pm40$		&	\\
\hline
0.179	&	$75\pm4$		&	\\
0.199	&	$75\pm5$		&	\\
0.352	&	$83\pm14$		&	\\
0.593	&	$104\pm13$		&	\citet{Moresco:2012} \\
0.68	&	$92\pm8$		&	\\
0.781	&	$105\pm12$		&	\\
0.875	&	$125\pm17$		&	\\
1.037	&	$154\pm20$		&	\\
\hline
0.07	&	$69\pm19.6$		&	\\
0.12	&	$68.6\pm26.2$	&	\citet{Zhang:2014} \\
0.2		&	$72.9\pm29.6$	&	\\
0.28	&	$88.8\pm36.6$	&	\\
\hline
1.363	&	$160\pm33.6$	&	\citet{Moresco:2015} \\
1.965	&	$186.5\pm50.4$	&	\\
\hline
0.3802	&	$83\pm13.5$		&	\\
0.4004	&	$77\pm10.2$		&	\\
0.4247	&	$87.1\pm11.2$	&	\citet{Moresco:2016} \\
0.4497	&	$92.8\pm12.9$	&	\\
0.4783	&	$80.9\pm9$		&	\\
\hline
\end{tabular}
\end{table}

In our analysis, we constrain the parameters by comparing the distance modulus from the Hubble parameter $H(z)$ and that from SNe Ia. However, there is no corresponding $H(z)$ data on the redshift of each observed SNe Ia data, which means the redshifts of SNe Ia and $H(z)$ are not one-to-one. Therefore, we should reconstruct a function $H(z)$ to ensure that each SNe Ia has corresponding $H(z)$ data at the same redshift. We reconstruct the function $H(z)$ versus the redshift using GaPP\footnote{http://www.acgc.uct.ac.za/~seikel/GAPP/index.html}, which is a python package proposed by \citet{Seikel:2012a} to execute the model-independent method GP. This method has been widely used in recent works \citep{Bilicki:2012,Seikel:2012b,Shafieloo:2012,Seikel:2013,Busti:2014,Yahya:2014,Yang:2015,Cai:2016,Wei:2017,Yu:2016,Zhang:2016}. The reconstructed function $H(z)$ is a Gaussian distribution with a mean value and Gaussian error at each point $z$. The functions at the different points of $z$ and $\tilde{z}$ are related by a covariance function $f(z, \tilde{z})$, which depends only on a set of hyperparameters $\ell$ and $\sigma_{f}$. The characteristic length scale $\ell$ can be thought of as the distance one has to travel in the $z$-direction to get a significant change in $H(z)$, whereas the signal variance $\sigma_{f}$ denotes the typical change in the $H$-direction. Both $\ell$ and $\sigma_{f}$ will be optimized by the GP with the observational data sets. After our careful checks, we obtain the same optimized values of $\ell$ and $\sigma_{f}$ when reconstructing the function $H(z)$ no matter how the initial values for these two hyperparameters are set. This implies that the reconstructed function $H(z)$ is not dependent on the initial hyperparameter settings and that we can safely use the reconstructed function.

For the reconstructed function $H(z)$, one can obtain the Hubble constant $H_0$ by setting $z=0$. On the other hand, it is worth noting that previous works (such as \citet{Wei:2017}) found that the existence of the Hubble constant $H_0$ when reconstructing a function $H(z)$ using the GP, could affect the final constraint on the parameters. So, in order to compare this with the results of no prior of $H_0$ when reconstructing a function $H(z)$, following the treatment of \citet{Zhang:2016} and \citet{Wei:2017}, we adopt two recent measurements of $H_0$ to reconstruct a function $H(z)$ at the same time: $H_0=67.74\pm 0.46~\rm km~ s^{-1} ~Mpc^{-1}$ with $0.7\%$ uncertainty \citep{Ade:2016} and $H_0=73.24\pm 1.74~\rm km~ s^{-1} ~Mpc^{-1}$ with $2.4\%$ uncertainty \citep{Riess:2016}. Hence, we have three cases when reconstructing a function $H(z)$:
\begin{itemize}
	\item[(a)] with no prior of $H_0$;
	\item[(b)] with prior of $H_0=67.74\pm 0.46~\rm km~ s^{-1} ~Mpc^{-1}$; and
	\item[(c)] with prior of $H_0=73.24\pm 1.74~\rm km~ s^{-1} ~Mpc^{-1}$.
\end{itemize}
The prior of $H_0$ in case (b) is indeed model-dependent since it is estimated in the $\Lambda$CDM model using CMB, SNe Ia, BAO, and the local $H_0$ data. But, we just want to compare the final constraint on the parameters for cases (b) and (c) with that of case (a). So, case (b) is acceptable in our model-independent analysis.

\begin{figure*}
	\centering
	\includegraphics[width=0.8\textwidth]{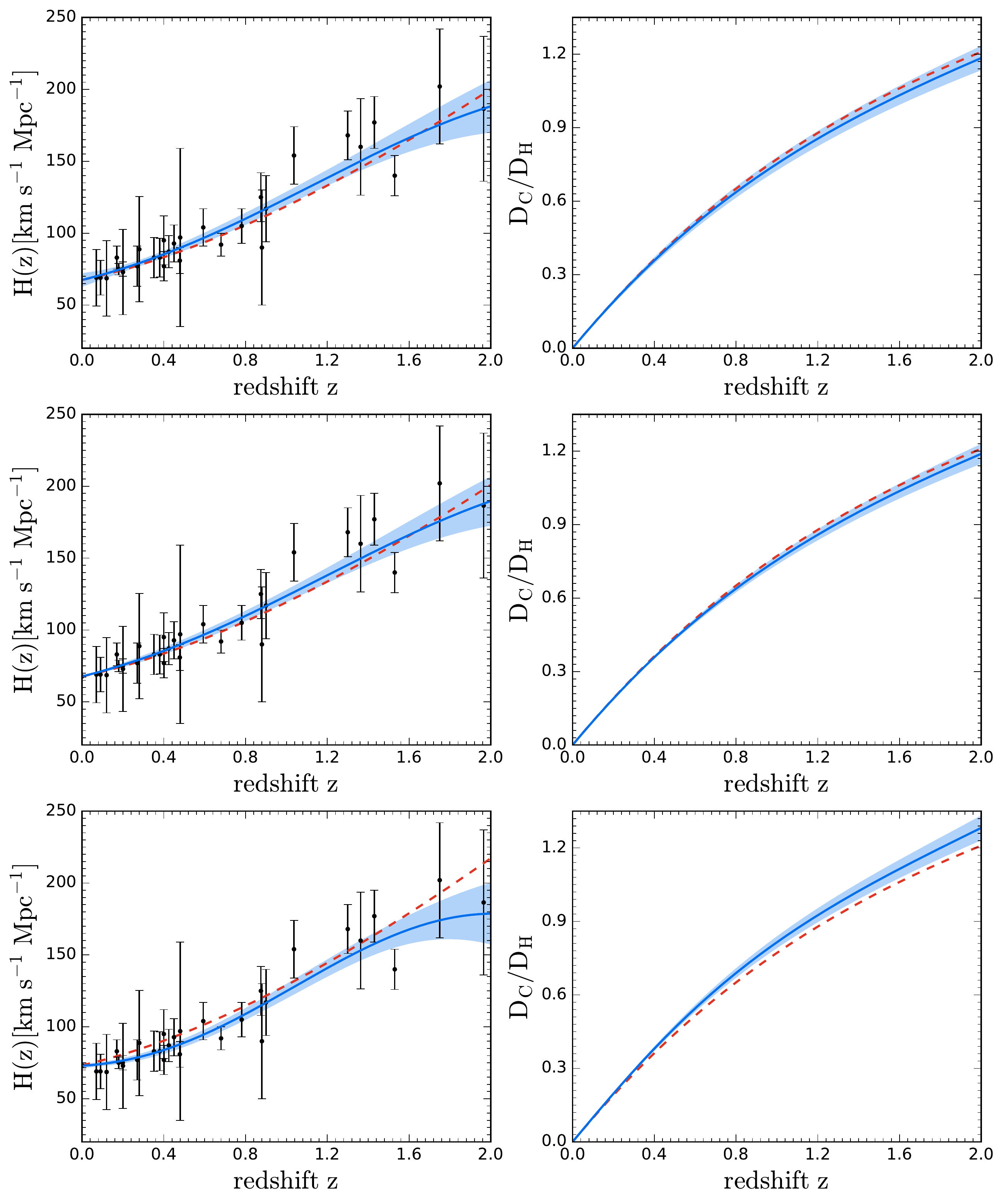}
	\caption{Reconstructed function of $H(z)$ and the corresponding reconstructed $D_C/D_{\rm H}$ by using the GP (blue lines). The blue regions are the $1\sigma$ confidence regions, while the red dashed lines correspond to the best-fit flat $\Lambda$CDM models with $\Omega_{\rm m}=0.3$. The black dots with error bars represent the CC $H(z)$ data. The top two panels correspond to the results of case (a), the two panels in the middle represent that of case (b), and the bottom two panels are for case (c). See the text for details.}\label{fig:rec_Hz}
\end{figure*}

The reconstructed functions of $H(z)$ are shown in Figure \ref{fig:rec_Hz}. The reconstructed local Hubble parameter with 68\% C.L. of case (a) is
\begin{equation}
H_0=67.56\pm 4.77~\rm km~ s^{-1} ~Mpc^{-1} ~,
\end{equation}
which is quite similar with the prior of $H_0=67.74\pm 0.46~\rm km~ s^{-1} ~Mpc^{-1}$. For comparison, we also plot the best-fit flat $\Lambda$CDM model with $\Omega_{\rm m}=0.3$ (red dashed lines of Figure \ref{fig:rec_Hz}). Note that $H_0$ is different for cases (a), (b), and (c), which means different $H_0$ for the best-fit flat $\Lambda$CDM model in the panels of Figure \ref{fig:rec_Hz}. In other words, $H_0=67.56~\rm km~ s^{-1} ~Mpc^{-1}, ~67.74~\rm km~ s^{-1} ~Mpc^{-1}$, and $73.24~\rm km~ s^{-1} ~Mpc^{-1}$ for cases (a), (b), and (c), respectively when plotting the best-fit flat $\Lambda$CDM model in Figure \ref{fig:rec_Hz}. It is obvious that the reconstructed functions are consistent with those of the flat $\Lambda$CDM model within a $1\sigma$ confidence level for the cases (a) and (b), with a little deviation for case (c).

Then we can get the total line-of-sight comoving distance $D_{C}$ \citep{Hogg:1999} by using
\begin{equation}\label{equ:comoving}
D_{C}=c\int_{0}^{z}\frac{dz'}{H(z')} ~.
\end{equation}
We obtain the error of $D_C$ by integrating the error of the function $H(z)$, which means that we integrate the upper and lower edge lines of the $1\sigma$ error regions in the left three panels of Figure \ref{fig:rec_Hz} by using Equation \ref{equ:comoving} to get the upper and lower edge lines of the $1\sigma$ error regions of $D_C$. The corresponding reconstructed $D_C/D_{\rm H}$ are shown in Figure \ref{fig:rec_Hz}.

The total line-of-sight comoving distance $D_{C}$ connects the luminosity distance $D_{L}$ via
\begin{equation}\label{equ:dl_Hz}
\frac{D_{L}}{(1+z)}=
\begin{cases}
\frac{D_{H}}{\sqrt{\Omega_K}} \sinh{[\sqrt{\Omega_K}D_{C}/D_{H}]}&\Omega_K>0\\D_{C} &\Omega_K=0\\ \frac{D_{H}}{\sqrt{\left| \Omega_K \right|}} \sin{[\sqrt{\left| \Omega_K \right|}D_{C}/D_{H}]}&\Omega_K<0 ~,
\end{cases}
\end{equation}
where $D_{H}=cH_0^{-1}$. So, we can get the luminosity distance $D_{L}(\Omega_K)$ from Equation (\ref{equ:dl_Hz}). Then, we can further obtain the reconstructed distance modulus of $H(z)$ by using
\begin{equation}\label{equ:mu_Hz}
\mu_{H}(\Omega_K)=5\log\frac{D_L(\Omega_K)}{\rm Mpc}+25 ~.
\end{equation}

\begin{figure}
	\centering
	\includegraphics[width=0.45\textwidth]{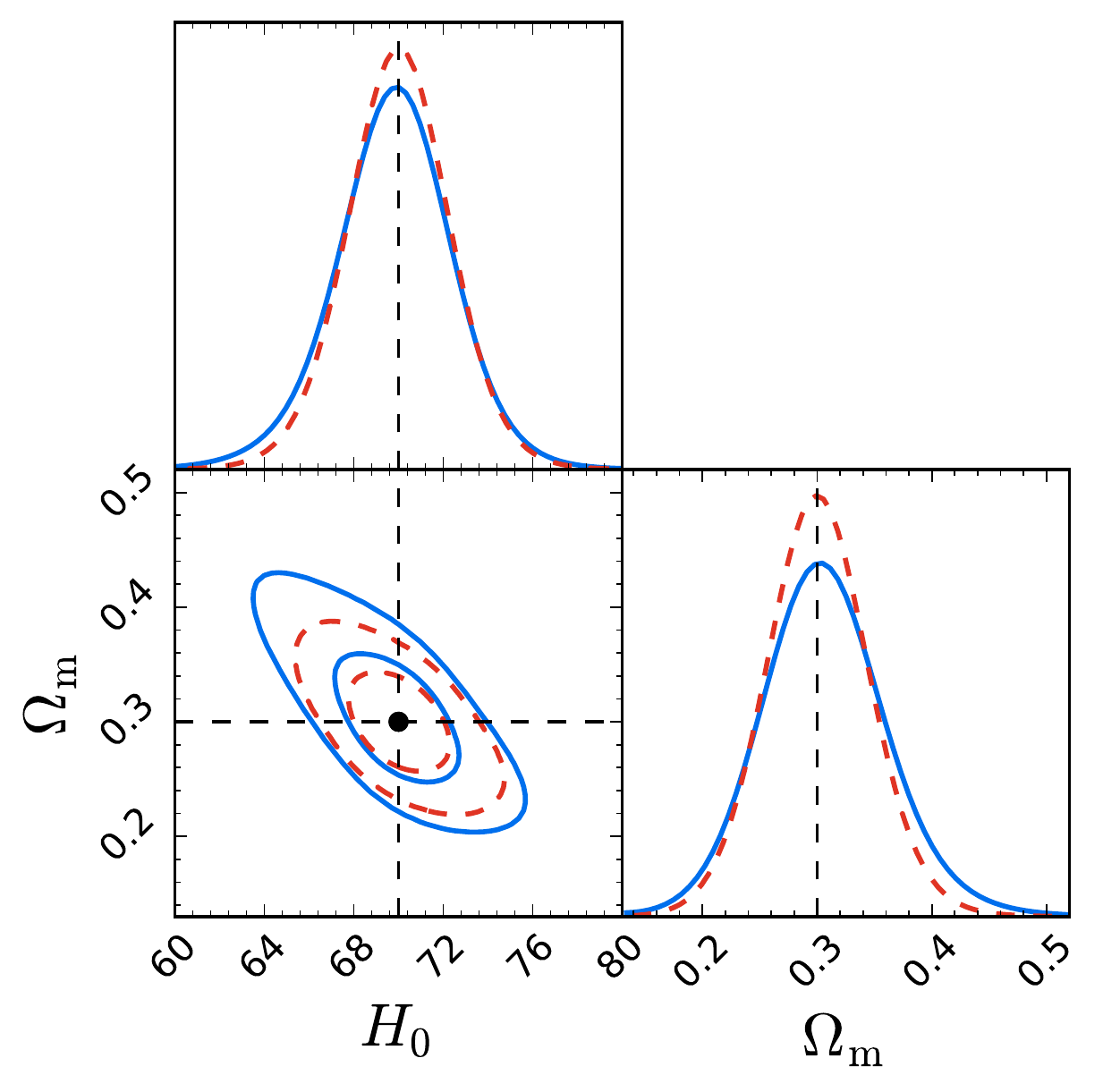}
	\caption{One-dimensional and two-dimensional marginalized distributions with $1\sigma$ and $2\sigma$ contours for $H_0$ and $\Omega_{\rm m}$. The black point represents the fiducial value of $H_0=70~\rm km~ s^{-1} ~Mpc^{-1}$ and $\Omega_{\rm m}=0.3$ when simulating the $H(z)$ data in the flat $\Lambda$CDM framework. The blue solid lines represent the results by comparing distance modulus, which have taken into account the GP method, while the red dashed lines are for that by comparing $H(z)$, which does not involve the GP method. See the text for details.}\label{fig:H0-omm}
\end{figure}

To check the reliability of the GP and the error processing method, we simulated the $H(z)$ data in the framework of a flat $\Lambda$CDM model by using
\begin{equation}
H(z) = H_0 \sqrt{\Omega_{\rm m}(1+z)^3 + 1-\Omega_{\rm m}}~.
\end{equation}
We set fiducial values at $H_0=70~\rm km~ s^{-1} ~Mpc^{-1}$ and $\Omega_{\rm m}=0.3$. It should be noted that $\Omega_K=0$ in the flat $\Lambda$CDM model. The uncertainties of the real $H(z)$ data have the mean relative error of 23.78\%. Therefore, we simulated the $H(z)$ data with the same uncertainty level. Based on the simulated $H(z)$ data, we use the GP method to reconstruct $\mu_{\rm H}$ and its corresponding error at a certain redshift of $z$. Then, we use the minimization function in Python to find the best-fit cosmological parameters that corresponds to minimum $\chi^2$.
\begin{equation}
\chi^2(H_0, \Omega_{\rm m}) = \sum_{i}\frac{\left[\mu_{\rm th}(z_i;H_0, \Omega_{\rm m})-\mu_{H}(z_i)\right]^{2}}{\sigma^2_{\mu_H,i}}~.
\end{equation}
We simulated 100,000 realizations of the data with different random seeds and repeated the minimization process. The marginalized two-dimensional constraint contour and one-dimensional probability density distributions of $H_0$ and $\Omega_{\rm m}$ are shown in Figure \ref{fig:H0-omm} (blue solid lines). Parameters with $1\sigma$ uncertainty are $H_0=69.924\pm1.511$ and $\Omega_{\rm m}=0.303\pm0.031$, which completely cover the fiducial values of the parameters (black point in Figure \ref{fig:H0-omm}) for the mock data. The recovered errors of $H_0$ and $\Omega_{\rm m}$ are similar to those obtained from a similar number of $H(z)$ data points in model-independent analyses and the GP method \citep{Busti:2014,Verde:2014}.

At the same time, we constrain $H_0$ and $\Omega_{\rm m}$ by comparing the simulated $H(z)$ with that of the flat $\Lambda$CDM model. This process does not involve the GP method, hence, the constraint on the parameters can be treated as the expected result. Figure \ref{fig:H0-omm} (red dashed lines) shows the constraint on $H_0$ and $\Omega_{\rm m}$. The best-fit values with $1\sigma$ errors of $H_0$ and $\Omega_{\rm m}$ are $H_0=70.002\pm1.091$ and $\Omega_{\rm m}=0.300\pm0.021$, which also completely cover the fiducial values of the parameters (black point in Figure \ref{fig:H0-omm}) for the mock data. One can find that the error bars of $H_0$ and $\Omega_{\rm m}$, which have taken into account the GP and error processing method, are similar to that of the expected precision. This indicates that the GP and the error processing method is reliable and unbiased, so it can be used for the subsequent analysis.

\subsection{SNe Ia data}\label{subsec:SNe_data}

In our analysis, we consider the latest 740 joint light curve (JLA) SNe Ia \citep{Betoule:2014}. The observed distance modulus of JLA SNe Ia is given by
\begin{equation}\label{equ:mu_jla_obs}
\mu_{JLA}^{obs}(\alpha, \beta, M_B, \Delta_M)=m_{B}^*-M_{B}+\alpha\times x_1-\beta\times c ~,
\end{equation}
where $m_{B}^*$ is the observed peak magnitude in the rest frame of the $B$ band and $M_B$ represents the absolute magnitude of the $B$ band, which is assumed to be related to the host stellar mass ($M_{\rm stellar}$) by a simple step function of
\begin{eqnarray}
M_{B}=\begin{cases} M_{B}^1~~~&\mathrm{if}~
M_{\mathrm{stellar}}<10^{10} M_\odot\\
M_{B}^1+\Delta_{M}~~&\mathrm{otherwise} ~,
\end{cases}
\end{eqnarray}
where $x_1$ is the time stretching of light curve, and $c$ corresponds to the supernova color at maximum brightness. $\alpha, \beta, M_B$ and $\Delta_M$ are nuisance parameters that should be constrained simultaneously with the cosmological parameters.

In particular, any effect that reduces the number of photons would dim the SNe Ia brightness and increase $D_L$. So, in order to use the full redshift range of the available data, we consider the following simple parameterization of a deviation from the CDD relation \citep{Avgoustidis:2009}:
\begin{equation}\label{equ:changed_CDDR}
D_{L,obs} = D_A(1+z)^{2+\epsilon} ~,
\end{equation}
where $\epsilon$ is parameterizing the opacity of the universe. Then, we obtain the $observed$ distance modulus using Equations (\ref{equ:CDDR}) and (\ref{equ:changed_CDDR}):
\begin{equation}\label{equ:3}
D_{L,obs} = D_{L,true}(1+z)^{\epsilon} ~.
\end{equation}
Hence, the $observed$ distance modulus is given by
\begin{eqnarray}\label{equ:mu_true}
\mu_{obs}(z) = \mu_{true}(z) + 5\epsilon\log(1+z) ~.
\end{eqnarray}
So, the $true$ distance modulus of JLA SNe Ia is
\begin{equation}\label{equ:mu_jla_true}
\mu_{JLA}^{true}(\epsilon,\alpha, \beta, M_B, \Delta_M)=\mu_{JLA}^{obs}(\alpha, \beta, M_B, \Delta_M)-5\epsilon\log(1+z) ~.
\end{equation}

Finally we fit $\Omega_K, \epsilon$, and other nuisance parameters simultaneously using the $H(z)$ and SNe Ia data by minimizing the $\chi^{2}$ statistic \citep{Betoule:2014,Wei:2017}:
\begin{equation}\label{equ:chi2_SNe}
\chi^{2}=\bf{\Delta \hat{\mu}}^{T}\cdot \textbf{Cov}^{-1}\cdot \bf{\Delta \hat{\mu}} ~,
\end{equation}
where $\Delta \hat{\mu}=\hat{\mu}_{JLA}^{true}(\epsilon,\alpha, \beta, M_B, \Delta_M)-\hat{\mu}_H(\Omega_K)$ is the difference between the distance moduli of JLA SNe Ia and that of the $H(z)$ data, and $\rm\bf Cov = \bar{D}_{\rm stat} + C_{\rm stat} + C_{\rm sys} $ is the full covariance matrix. Here $\rm\bf\bar{D}_{\rm stat}$ is the diagonal part of the statistical uncertainty:
\begin{equation}
\bf(\bar{D}_{\rm stat})_{\textit{ii}}=\bf (D_{\rm stat}^{\rm SNe})_{\textit{ii}} + \sigma^{\rm 2}_{\mu_{\textit{H,i}}} ~,
\end{equation}
where $\bf D_{\rm stat}^{\rm SNe}$ comes from SNe Ia, and $\sigma^{\rm 2}_{\mu_{\textit{H}}}$ comes from the $H(z)$ data. After a careful check, we find that the errors of JLA SNe Ia play a dominant role. Thus, the errors of $H(z)$ have little effect on the constraint on the parameters. The full covariance data have been used for the JLA data set. We refer the reader to \citet{Betoule:2014} and \citet{Wei:2017} for details about the method of calculation.

\section{Results}\label{sec:results}

\begin{figure}
	\centering
	\includegraphics[width=0.45\textwidth]{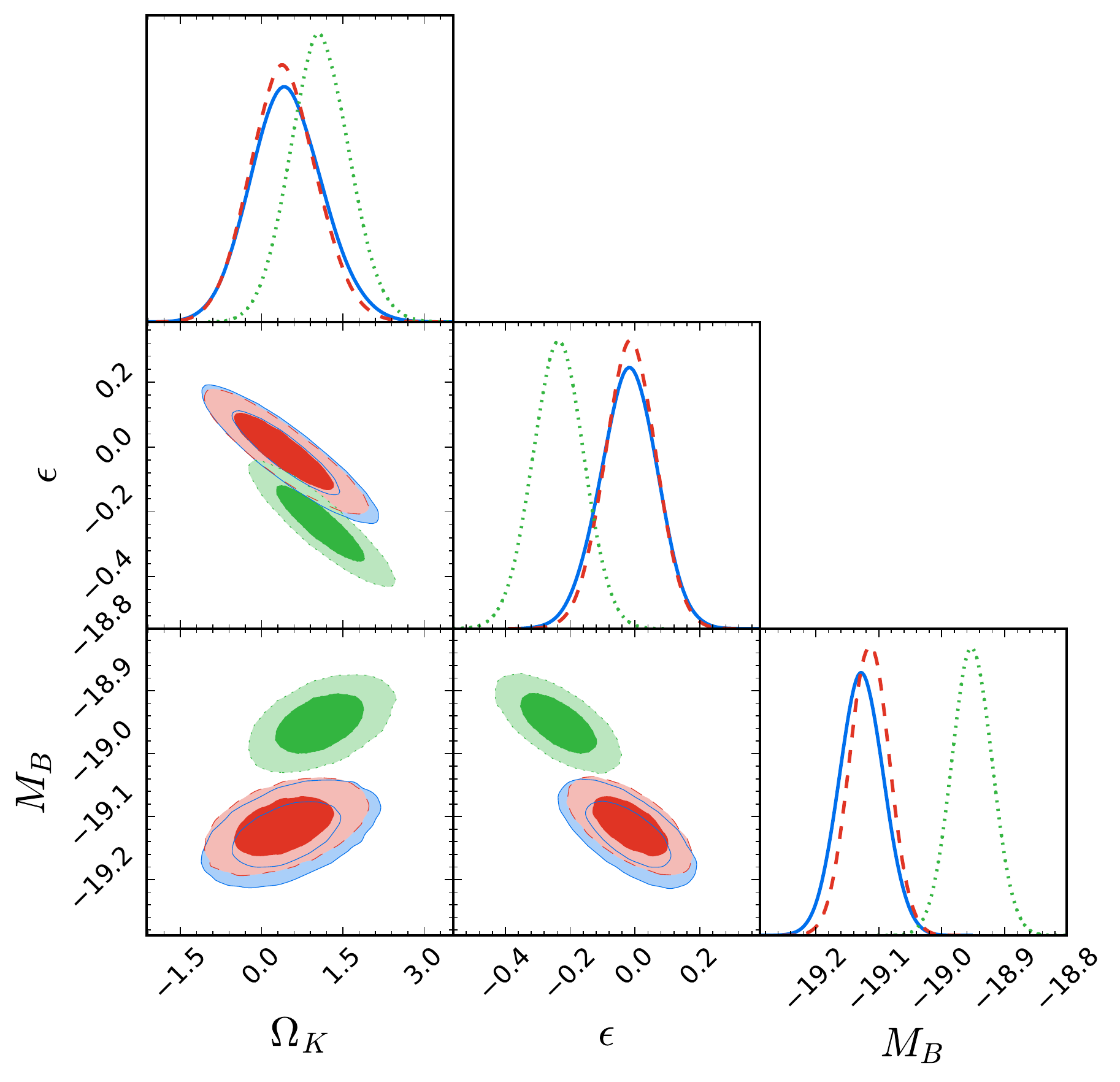}
	\caption{One-dimensional and two-dimensional marginalized distributions with $1\sigma$ and $2\sigma$ contours for $\Omega_K$, $\epsilon$, and $M_B$ using $H(z)$ + JLA. The blue solid lines, red dashed lines, and green dotted lines represent the results of case (a), (b), and (c), respectively. See the text for details.}\label{fig:params_union}
\end{figure}

We constrain $\Omega_K, \epsilon$, and other nuisance parameters simultaneously by comparing the distance modulus taken from the reconstructed function $H(z)$ and that from JLA by minimizing $\chi^2$ (Equation \ref{equ:chi2_SNe}). We use {\it emcee}\footnote{https://pypi.python.org/pypi/emcee}~introduced by \citet{Foreman-Mackey:2013}, a Python module that uses the Markov Chain Monte Carlo method to get the best-fit values and their uncertainties of parameters by generating sample points of the probability distribution, to constrain $\Omega_K, \epsilon$, and other nuisance parameters simultaneously. In constraining parameters, the Hubble constant $H_0$ is set to a value based on the reconstructed function $H(z)$. In other words, the values of $H_0$ are
\begin{itemize}
	\item[$H_0$] $=67.56\pm 4.77~\rm km~ s^{-1} ~Mpc^{-1}$ for case (a),
	\item[$H_0$] $=67.74\pm 0.46~\rm km~ s^{-1} ~Mpc^{-1}$ for case (b),
	\item[$H_0$] $=73.24\pm 1.74~\rm km~ s^{-1} ~Mpc^{-1}$ for case (c),
\end{itemize}
when constraining $\Omega_K, \epsilon$, and other nuisance parameters simultaneously. The best-fit values with $1\sigma$ errors of $\Omega_K$ and $\epsilon$ are shown in Table \ref{tab:omk_epsilon}. We plot the constraints on $\Omega_K$, $\epsilon$, and $M_B$ in Figure \ref{fig:params_union}. The 68\% C.L. constraint on $\Omega_K$ for case (a) is
\begin{equation}
\Omega_K=0.440\pm0.645 ~, 
\end{equation}
which is consistent with the latest constraint $\Omega_K=0.09\pm0.25$ \citep{Wei:2017} within a 1$\sigma$ confidence level. Meanwhile, the 86\% C.L. constraint on cosmic opacity $\epsilon$ is
\begin{equation}
\epsilon=-0.018\pm0.084 ~,
\end{equation}
which is also consistent with $\epsilon = -0.04^{+0.08}_{-0.07} (2\sigma)$ \citep{Avgoustidis:2010} and $\epsilon=0.017\pm0.052 (1\sigma)$ \citep{Holanda:2013} within a $1\sigma$ confidence level.

Figure \ref{fig:params_union} shows that the best-fit values are not consistent with each other for the three cases of the reconstructed $H(z)$. This means that the constraints on cosmic curvature parameter $\Omega_K$, opacity parameter $\epsilon$, and the absolute magnitude $M_B$ of SNe Ia are influenced by the prior $H_0$ when reconstructing the $H(z)$ data. In other words, $H_0$ has correlations with $\Omega_K$, $\epsilon$, and $M_B$. This is also found by \citet{Wei:2017}. Figure 　\ref{fig:params_union} also shows that $\Omega_K$ and $\epsilon$ are not independent of $M_B$, which means that the light curve parameters of SNe Ia should be set free when fitting $\Omega_K$ and $\epsilon$. On the other hand, it is worth noting that there are strong correlations between $\Omega_K$ and $\epsilon$ for all of the results. It is not difficult to see that a fine tuning of $\epsilon$ will cause a big change for $\Omega_K$, and in turn, one can get different values of $\epsilon$ when fine tuning $\Omega_K$. This implies that the cosmic opacity should be considered beside the light curve parameters of SNe Ia when fitting the cosmic curvature parameter $\Omega_K$.

\begin{table}
	\centering\caption{$1\sigma$ Constraints on $\Omega_K$ and $\epsilon$ for $H(z)$+JLA. Case (a), (b) and (c) represent no prior of $H_0$, $H_0 = 67.74\pm 0.46~\rm km~ s^{-1} ~Mpc^{-1}$, and $H_0 = 73.24\pm 1.74~\rm km~ s^{-1} ~Mpc^{-1}$ when reconstructing a function $H(z)$, respectively. See the text for details.}\label{tab:omk_epsilon}
	\begin{tabular}{c|c|c|c}
		\hline\hline
		Cases & (a) & (b) & (c)  \\
		\hline
		$\Omega_K$ & $0.440\pm0.645$ & $0.374\pm0.580$ & $1.044\pm0.514$  \\
		$\epsilon$ & $-0.018\pm0.084$ & $-0.012\pm0.075$ & $-0.232\pm0.075$  \\
		\hline\hline
	\end{tabular}
\end{table}

Interestingly, one can see that the results of case (a) are very similar to that of case (b), while there is a big difference between cases (a) and (c) for their constraint on the parameters. After careful inspection, we find that the reason for this phenomenon is the prior $H_0$ when reconstructing the $H(z)$ data. Both the best-fit value and its error of the prior $H_0$ will affect the constraint on parameters, in which the best-fit value of the prior $H_0$ is the main factor. Note that the best-fit value of the reconstructed local Hubble parameter for case (a) is $H_0=67.56~\rm km~ s^{-1} ~Mpc^{-1}$, which is very similar to the prior of $H_0=67.74~\rm km~ s^{-1} ~Mpc^{-1}$ for case (b), and is very different from the prior of $H_0=73.24~\rm km~ s^{-1} ~Mpc^{-1}$ for case (c). So, it is easy to understand why the results of case (a) are very similar to that of case (b), and the results of case (c) are very different from those of the other two cases. Moreover, Table \ref{tab:omk_epsilon} shows that the current observations are compatible with a flat and transparent universe within a $1\sigma$ confidence level for cases (a) and (b). For the case (c), the constraints on $\Omega_K$ and $\epsilon$ are
\begin{align}
\nonumber \Omega_K&=1.044\pm 1.070(2\sigma)\pm 1.638(3\sigma)\\
\epsilon&=-0.232\pm 0.155(2\sigma)\pm 0.237(3\sigma) ~,
\end{align}
which means that a flat and transparent universe is also supported by case (c) within a 3$\sigma$ confidence level.

\section{Conclusions and Discussions}\label{sec:conclusions_discussions}

In this paper, on the basis of \citet{Li:2016} and \citet{Wei:2017}, we test the cosmic curvature and the opacity simultaneously by combining the Hubble parameter $H(z)$ with the SNe Ia data. First, we reconstruct a function of $H(z)$ using the GP, which is a model-independent method. Then we obtain the corresponding reconstructed distance modulus $\mu_{\rm H}$. Two priors of $H_0$ ($67.74\pm 0.46~\rm km~ s^{-1} ~Mpc^{-1}$ and $73.24\pm 1.74~\rm km~ s^{-1} ~Mpc^{-1}$) are considered when reconstructing the $H(z)$ data to investigate the influence of $H_0$ on the constraints on the parameters. In addition, the light curve fitting parameters that account for the distance estimation of SNe Ia are set free to test whether $\Omega_K$ and $\epsilon$ are dependent on them. Finally, we constrain the cosmic curvature parameter $\Omega_K$, cosmic opacity parameter $\epsilon$, and other nuisance parameters simultaneously by comparing $\mu_{\rm H}$ with the distance modulus of SNe Ia.

We use 30 $H(z)$ and 740 JLA SNe Ia to execute our analysis. The results show that the constraints on $\Omega_K$, $\epsilon$, and $M_B$ are sensitive to the prior of $H_0$ when reconstructing a function of $H(z)$. Both $\Omega_K$ and $\epsilon$ have a dependence on the light curve fitting parameter $M_B$ of SNe Ia. In addition, we find that $\Omega_K$ is strongly degenerate with $\epsilon$, which leads to a fine tuning of $\epsilon$, which will cause a big change of $\Omega_K$. Hence, the cosmic opacity should be considered when fitting $\Omega_K$. Moreover, the results show that a flat and transparent universe is supported by the current observational data.

\section*{Acknowledgements}
We thank Kai Liao and Xuheng Ding for helpful discussions. G.-J.W. and Z.-H.Z. are supported by the NSFC under grant No. 11633001. J.-J.W. is supported by the NSFC under grant No. 11603076, the Natural Science Foundation of Jiangsu Province (BK20161096), and the Youth Innovation Promotion Association CAS (2017366). Z.-X.L. is supported by the NSFC under grant No. 11505008. J.-Q.X. is supported by the National Youth Thousand Talents Program and the NSFC under grant Nos. 11422323 and 11690023. The research is also supported by the Fundamental Research Funds for the Central Universities Nos. 2017EYT01 and 2017STUD01.


\begin{thebibliography}{}
\bibitem[\protect\citeauthoryear{Ade et al.}{2016}]{Ade:2016}Ade, P. A. R., Aghanim, N., Arnaud, M., et al. 2016, A\&A, 594, A13 

\bibitem[\protect\citeauthoryear{Amanullah et al.}{2010}]{Amanullah:2010}Amanullah, R., Lidman, C., Rubin, D., et al. 2010, ApJ, 716, 712 

\bibitem[\protect\citeauthoryear{Avgoustidis et al.}{2009}]{Avgoustidis:2009}Avgoustidis, A., Verde, L. \& Jimenez, R., 2009, JCAP 0906, 012

\bibitem[\protect\citeauthoryear{Avgoustidis et al.}{2010}]{Avgoustidis:2010}Avgoustidis, A., Burrage, C., Redondo, J., Verde, L., \& Jimenez, R., 2010, JCAP, 1010, 024

\bibitem[\protect\citeauthoryear{Bassett \& Kunz}{2004}]{Bassett:2004}Bassett, B. A. \& Kunz, M., 2004, Phys. Rev. D, 69, 101305

\bibitem[\protect\citeauthoryear{Bernstein}{2006}]{Bernstein:2006}Bernstein, G. 2006, ApJ, 637, 598

\bibitem[\protect\citeauthoryear{Betoule et al.}{2014}]{Betoule:2014}Betoule, M., Kessler, R., Guy, J., et al. 2014, A\&A, 568, A22

\bibitem[\protect\citeauthoryear{Bilicki \& Seikel}{2012}]{Bilicki:2012}Bilicki, M., \& Seikel, M. 2012, MNRAS, 425, 1664

\bibitem[\protect\citeauthoryear{Blake et al.}{2012}]{Blake:2012}Blake, C., Brough, S., Colless, M., et al. 2012, MNRAS, 425, 405

\bibitem[\protect\citeauthoryear{Bolton et al.}{2008}]{Bolton:2008}Bolton, A. S., Burles, S., Koopmans, L. V. E., et al. 2008, ApJ, 682, 964

\bibitem[\protect\citeauthoryear{Busti et al.}{2014}]{Busti:2014}Busti, V. C., Clarkson, C., \& Seikel, M. 2014, MNRAS, 441, L11

\bibitem[\protect\citeauthoryear{Cai et al.}{2016}]{Cai:2016}Cai, R.-G., Guo, Z.-K., \& Yang, T., \ 2016, PRD, 93, 043517

\bibitem[\protect\citeauthoryear{Chen et al.}{2012}]{Chen:2012}Chen, J., Wu, P., Yu, H., \& Li, Z. 2012, JCAP, 10, 029

\bibitem[\protect\citeauthoryear{Chuang \& Wang}{2012}]{Chuang:2012}Chuang, C.-H., \& Wang, Y. 2012, MNRAS, 426, 226

\bibitem[\protect\citeauthoryear{Clarkson et al.}{2007}]{Clarkson:2007}Clarkson, C., Cortes, M., \& Bassett, B., JCAP, 2007, 08, 011

\bibitem[\protect\citeauthoryear{Clarkson et al.}{2008}]{Clarkson:2008}Clarkson, C., Bassett, B. A., \& Hui-Ching Lu, T., 2008, PRL, 101, 011301

\bibitem[\protect\citeauthoryear{Daly \& Djorgovski}{2003}]{Daly:2003}Daly, R. A. \& Djorgovski, S. G., 2003, Astrophys. J. 597, 9

\bibitem[\protect\citeauthoryear{Ellis}{2007}]{Ellis:2007}Ellis G. F. R. , Gen. Relativ. Gravit. 39, 1047 (2007).

\bibitem[\protect\citeauthoryear{Etherington}{1993}]{Etherington:1993}Etherington I. M. H., Philos. Mag. 15, 761 (1933).

\bibitem[\protect\citeauthoryear{Fu \& Li}{2017}]{Fu:2017}Fu, X. \& Li, P., 2017, arXiv: 1702.03626

\bibitem[\protect\citeauthoryear{Foreman-Mackey et al.}{2013}]{Foreman-Mackey:2013}Foreman-Mackey, D., Hogg, D. W., Lang, D. and Goodman, J. 2013, PASP, 125, 306F

\bibitem[\protect\citeauthoryear{Gaztanaga et al.}{2009}]{Gaztanaga:2009}Gazta\~{n}aga, E., Cabr\'{e}, A., \& Hui, L. 2009, MNRAS, 399, 1663

\bibitem[\protect\citeauthoryear{Gurvits}{1994}]{Gurvits:1994}Gurvits, L. I., 1994, Astrophys. J., 425, 442

\bibitem[\protect\citeauthoryear{Gonçalves et al.}{2012}]{Gonçalves:2012}Gonçalves, R. S., Holanda, R. F. L., \& Alcaniz, J. S., 2012, Mon. Not. R. Astron. Soc., 420, L43

\bibitem[\protect\citeauthoryear{Hogg}{1999}]{Hogg:1999}Hogg, D. W. 1999, ArXiv Astrophysics e-prints, astro-ph/9905116

\bibitem[\protect\citeauthoryear{Holanda et al.}{2010}]{Holanda:2010}Holanda, R. F. L., Lima, J. A. S., \& Ribeiro, M. B., 2010, Astrophys. J., 722, L233

\bibitem[\protect\citeauthoryear{Holanda et al.}{2011}]{Holanda:2011}Holanda, R.~F.~L., Lima, J.~A.~S. \& Ribeiro, M.~B., 2011, \aap, 528, 14

\bibitem[\protect\citeauthoryear{Holanda et al.}{2012a}]{Holanda:2012a}Holanda, R. F. L., Lima, J. A. S., \& Ribeiro, M. B., 2012, Astron. Astrophys., 538, A131

\bibitem[\protect\citeauthoryear{Holanda et al.}{2012b}]{Holanda:2012b}Holanda, R. F. L., Gonçalves, R. S., Alcaniz, J. S.,  2012, J. Cosmol. Astropart. Phys., 06, 022.

\bibitem[\protect\citeauthoryear{Holanda et al.}{2013}]{Holanda:2013}Holanda, R. F. L., Carvalho, J. C., \& Alcaniz, J. S., 2013, J. Cosmol. Astropart. Phys., 04, 027

\bibitem[\protect\citeauthoryear{Holanda \& Busti}{2014}]{Holanda:2014}Holanda, R. F. L. \& Busti, V. C., 2014, Phys. Rev. D, 89, 103517

\bibitem[\protect\citeauthoryear{Hu et al}{2017}]{Hu:2017}Hu, J., Yu, H. \& Wang, F.~Y., 2017, ApJ, 836, 107

\bibitem[\protect\citeauthoryear{Jesus et al.}{2016}]{Jesus:2016}Jesus, J.~F., {Holanda}, R.~F.~L. \& {Dantas}, M.~A., 2016, arXiv: 1605.01342

\bibitem[\protect\citeauthoryear{Jimenez \& Loeb}{2002}]{Jimenez:2002}Jimenez, R., \& Loeb, A. 2002, ApJ, 573, 37

\bibitem[\protect\citeauthoryear{Jimenez et al}{2003}]{Jimenez:2003}Jimenez, R., Verde, L., Treu, T., \& Stern, D. 2003, ApJ, 593, 622

\bibitem[\protect\citeauthoryear{Komatsu et al.}{2011}]{Komatsu:2011}Komatsu, E., Smith, K. M., Dunkley, J., et al. 2011, ApJS, 192, 18

\bibitem[\protect\citeauthoryear{Li et al.}{2013}]{Li:2013}Li, Z., Wu, P., Yu, H., \& Zhu, Z.-H., 2013, Phys. Rev. D 87, 103013

\bibitem[\protect\citeauthoryear{Li et al.}{2014}]{Li:2014}Li, Y.-L., Li, S.-Y., Zhang, T.-J., \& Li, T.-P. 2014, ApJL, 789, L15

\bibitem[\protect\citeauthoryear{Li et al.}{2016}]{Li:2016}Li, Z., Wang, G.-J., Liao, K., \& Zhu, Z.-H. 2016, ApJ, 833, 240

\bibitem[\protect\citeauthoryear{Liao et al.}{2013}]{Liao:2013}Liao, K. and Li, Z. and Ming, J. and Zhu, Z.-H., 2013, PhLB, 718, 1166L

\bibitem[\protect\citeauthoryear{Liao et al.}{2015}]{Liao:2015}Liao, K., Avgoustidis, A. \& Li, Z., 2015, Phys. Rev. D, 92, 123539

\bibitem[\protect\citeauthoryear{Lima et al.}{2011}]{Lima:2011}Lima, J. A. S., Cunha, J. V., \& Zanchin, V. T., 2011, Astrophys. J., 742, L26

\bibitem[\protect\citeauthoryear{Lv \& Xia}{2016}]{Lv:2016}Lv, M.-Z. \& Xia, J.-Q., 2016, Physics of the Dark Universe, 13, 139

\bibitem[\protect\citeauthoryear{More et al.}{2016}]{More:2016}More, S., Niikura, H., Schneider, J., Schuller, F.~P. \& Werner, M.~C., 2016, ArXiv e-prints, arXiv: 1612.08784

\bibitem[\protect\citeauthoryear{Moresco et al.}{2012}]{Moresco:2012}Moresco, M., Verde, L., Pozzetti, L., Jimenez, R., \& Cimatti, A. 2012, JCAP, 7, 053

\bibitem[\protect\citeauthoryear{Moresco et al.}{2015}]{Moresco:2015}Moresco, M. 2015, MNRAS, 450, L16

\bibitem[\protect\citeauthoryear{Moresco et al.}{2016}]{Moresco:2016}Moresco, M., Pozzetti, L., Cimatti, A., et al. 2016, JCAP, 5, 014

\bibitem[\protect\citeauthoryear{M\"{o}rtsell \& J\"{o}nsson}{2011}]{Mortsell:2011}M\"{o}rtsell, E., \& J\"{o}nsson, J., arXiv:1102.4485

\bibitem[\protect\citeauthoryear{Perlmutter et al.}{1999}]{Perlmutter:1999}Perlmutter, S., Aldering, G., Goldhaber, G., et al. 1999, ApJ, 517, 565 

\bibitem[\protect\citeauthoryear{Rana et al.}{2016}]{Rana:2016}Rana, A., Jain, D., Mahajan, S., Mukherjee, A., 2016, JCAP, 07, 026

\bibitem[\protect\citeauthoryear{Rana et al.}{2017}]{Rana:2017}Rana, A., Jain, D., Mahajan, S. \& Mukherjee, A. 2017, JCAP, 03, 028R

\bibitem[\protect\citeauthoryear{R\"{a}s\"{a}nen et al.}{2015}]{Rasanen:2015}R\"{a}s\"{a}nen, S., Bolejko, K., \& Finoguenov, A., \ 2015, PRL, 115, 101301

\bibitem[\protect\citeauthoryear{Riess et al.}{1998}]{Riess:1998}Riess, A. G., Filippenko, A. V., Challis, P., et al. 1998, AJ, 116, 1009 

\bibitem[\protect\citeauthoryear{Riess et al.}{2016}]{Riess:2016}Riess, A. G., Macri, L. M., Hoffmann, S. L., et al. 2016, ApJ, 826, 56

\bibitem[\protect\citeauthoryear{Samushia et al.}{2013}]{Samushia:2013}Samushia, L., Reid, B. A., White, M., et al. 2013, MNRAS, 429, 1514

\bibitem[\protect\citeauthoryear{Sapone et al.}{2014}]{Sapone:2014}Sapone, D.,  Majerotto, E., \& Nesseris, S., \ 2014, PRD, 90, 023012

\bibitem[\protect\citeauthoryear{Seikel et al.}{2012a}]{Seikel:2012a} Seikel, M., Clarkson, C., \& Smith, M.\ 2012, jcap, 6, 036 

\bibitem[\protect\citeauthoryear{Seikel et al.}{2012b}]{Seikel:2012b}Seikel, M., Yahya, S., Maartens, R., \& Clarkson, C. 2012b, Phys. Rev. D, 86, 083001

\bibitem[\protect\citeauthoryear{Seikel \& Clarkson}{2013}]{Seikel:2013}Seikel, M., \& Clarkson, C. 2013, ArXiv e-prints, arXiv:1311.6678

\bibitem[\protect\citeauthoryear{Shafieloo \& Clarkson}{2010}]{Shafieloo:2010}Shafieloo, A. \& Clarkson, C.  PRD, 81, 083537

\bibitem[\protect\citeauthoryear{Shafieloo et al.}{2012}]{Shafieloo:2012}Shafieloo, A., Kim, A. G., \& Linder, E. V. 2012, Phys. Rev. D, 85, 123530

\bibitem[\protect\citeauthoryear{Simon et al.}{2005}]{Simon:2005}Simon, J., Verde, L., \& Jimenez, R. 2005, Phys. Rev. D, 71, 123001

\bibitem[\protect\citeauthoryear{Stern et al.}{2010}]{Stern:2010}Stern, D., Jimenez, R., Verde, L., Kamionkowski, M., \& Stanford, S. A. 2010, JCAP, 2, 008

\bibitem[\protect\citeauthoryear{Sullivan et al.}{2010}]{Sullivan:2010}Sullivan, M., Conley, A., Howell, D. A., et al. 2010, MNRAS, 406, 782

\bibitem[\protect\citeauthoryear{Suzuki et al.}{2012}]{Suzuki:2012}Suzuki, N., Rubin, D., Lidman, C., et al. 2012, ApJ, 746, 85

\bibitem[\protect\citeauthoryear{Verde et al.}{2014}]{Verde:2014}Verde, L., Protopapas, P., Jimenez, R., 2014, PDU, 5, 307V

\bibitem[\protect\citeauthoryear{Wei \& Wu}{2017}]{Wei:2017}Wei, J.-J. \& Wu, X.-F., 2017, ApJ, 838, 160w

\bibitem[\protect\citeauthoryear{Yahya et al.}{2014}]{Yahya:2014}Yahya, S., Seikel, M., Clarkson, C., Maartens, R., \& Smith, M. 2014, PhRvD, 89, 023503

\bibitem[\protect\citeauthoryear{Yang et al.}{2015}]{Yang:2015}Yang, T., Guo, Z.-K., \& Cai, R.-G. 2015, Phys. Rev. D, 91, 123533

\bibitem[\protect\citeauthoryear{Yu \& Wang}{2016}]{Yu:2016}Yu, H., \& Wang, F. Y. 2016, ApJ, 828, 85

\bibitem[\protect\citeauthoryear{Zhang et al}{2014}]{Zhang:2014}Zhang, C., Zhang, H., Yuan, S., et al. 2014, Research in Astronomy and Astrophysics, 14, 1221

\bibitem[\protect\citeauthoryear{Zhang \& Xia}{2016}]{Zhang:2016}Zhang, M.-J., \& Xia, J.-Q., 2016, JCAP, 12, 005Z

\end{thebibliography}
\end{document}